\documentclass[preprint,groupedaddress,twocolumn,letterpaper,tightenlines,balancelastpage,lengthcheck,prl,10pt]{revtex4}
\usepackage{amsmath}
\usepackage{graphicx}

\begin{document}

\title{Chirality-- and thickness-dependent thermal conductivity of few-layer
graphene: a molecular dynamics study}
\author{Wei-Rong Zhong$^{1}$, Mao-Ping Zhang$^{1}$, Bao-Quan Ai$^{2}$}
\email{wrzhong@hotmail.com, aibq@scnu.edu.cn}
\author{Dong-Qin Zheng$^{1}$}
\affiliation{$^{1}$\textit{Department of Physics and Siyuan Laboratory, College of
Science and Engineering, Jinan University, Guangzhou 510632, P. R. China}}
\affiliation{$^{2}$\textit{Laboratory of Quantum Information Technology, ICMP and SPTE,
South China Normal University, Guangzhou, 510006 P. R. China}}
\date{\today}

\begin{abstract}
The thermal conductivity of graphene nanoribbons (layer from 1 to 8 atomic
planes) is investigated by using the nonequilibrium molecular dynamics
method. We present that the room-temperature thermal conductivity decays
monotonically with the number of the layers in few-layer graphene. The
superiority of zigzag graphene in thermal conductivity is only available in
high temperature region and disappears in multi-layer case. It is explained
that the phonon spectral shrink in high frequency induces the change of
thermal conductivity. It is also reported that single-layer graphene has
better ballistic transport property than the multi-layer graphene.
\end{abstract}

\keywords{Thermal conductivity, graphene, carbon, molecular dynamics
simulation}
\pacs{65.80.Ck\qquad Thermal properties of graphene }
\pacs{81.05.ue Graphene}
\maketitle

%\pacs{65.80.+n Thermal properties of small particles, nanocrystals,
%nanotubes }
%\pacs{ 81.07.De Nanotubes }

In the past decade, more and more attentions have been given to the question
of what happens with thermal conductivity when goes to low-dimensional
materials \cite{Lepri}. A two-dimensional materials-graphene \cite%
{Novoselov1}, in addition to its exceptional electric \cite{YBZhang} and
optical properties \cite{Nair}$^{,}$ \cite{Balandin}, reveals unique high
thermal conductivity. Thermal conductivity of single-layer graphene as well
as of carbon nanotubes is dependent on the chirality \cite{WZhang}. Recent
theoretical studies suggest that the thermal conductivity of single-layer
zigzag graphene is 20-50\% larger than that of the single-layer armchair
graphene \cite{Jiuning}. However, whether the superior thermal conductivity
of zigzag graphene remains available for multi-layer graphene has not got
enough attention and concern.

Additionally, experimental demonstrations have shown that the thermal
conductivity gets a decrease at the two- to three-dimensional (2D to 3D)
crossover of few-layer graphene \cite{Ghosh}. The fact that the thermal
conductivity of large enough graphene sheets should be higher than that of
basal planes of bulk graphite was predicted theoretically by Klemens \cite%
{Klemens}. Generally, thermal transport in conventional thin films still
retains `bulk' features because the cross-sections of these structures are
measured in many atomic layers. Heat conduction in such nanostructures is
dominated by extrinsic effects, for example, phonon-boundary or
phonon-defect scattering \cite{Hochbaum}. A recent experimental observation
of high-quality few-layer graphene materials shows that the room-temperature
thermal conductivity changes from \symbol{126}2,800 to \symbol{126}1,300 Wm$%
^{-1}$K$^{-1}$ when the number of atomic planes in few-layer graphene
increases from 2 to 4. It is explained that the observed evolution from two
dimensions to bulk attributed to the cross-plane coupling of the low-energy
phonons and changes in the phonon Umklapp scattering \cite{Ghosh}.

Recently, the method of molecular dynamics simulation has been successful in
discovering thermal conductivity and thermal rectification of the
nanostructures \cite{Jiuning}$^{,}$ \cite{GWu1}. This method, which builds
the system from the bottom up, is useful to understand the intrinsic
behavior, i.e., the phonon spectral behavior behind the significant change
of a material's ability to conduct heat \cite{Savin}. In this paper, we will
study the thermal conductivity of graphene ribbons (layer from 1 to 8 atomic
planes) by using the nonequilibrium molecular dynamics method. By
investigating the thermal conductivity and the phonon spectral behavior of
multi-layer graphene, we try to reveal the coupling of the low-energy
phonons from a more fundamental principle. The obtained results are of
significance for understanding the thermal properties in low-dimensional
materials and may open up few-layer graphene applications in the nanoscale
thermal device such as thermal diode \cite{Chang}, thermal transistor \cite%
{BWLi} and so on.

In our simulations, we have used classical molecular dynamics method based
on the Tersoff-Brenner potential \cite{Tersoff} of C-C bonding interactions.
The equations of motion for atoms in either the left or right Nos\'{e}%
-Hoover thermostat are \cite{Jiuning}$^{,}$ \cite{GWu2}%
\begin{equation}
\frac{d}{dt}p_{i}=F_{i}-\Gamma p_{i};\frac{d}{dt}\Gamma =\frac{1}{Q}\left[
\underset{i}{\sum }\frac{p_{i}^{2}}{2m_{i}}-3Nk_{B}T_{0}/2\right] ,
\end{equation}%
where $p_{i}$ is the momentum and $F_{i}$ is the force applied on the $i$-th
atom. $Q=3Nk_{B}T_{0}\tau ^{2}/2$, where $\tau $ is the relaxation time,
which is kept as $1ps$. $\Gamma $ is the dynamic parameter of the
thermostat, $T(t)$,\ which is defined as $\frac{m_{i}}{3k_{B}}%
(vx(t)^{2}+vy(t)^{2}+vz(t)^{2})$, where $v(t)$ is the time-dependent
velocity, is the instant temperature of the heat baths at time $t$. $T_{0}$ (%
$T_{L}$ or $T_{R}$). The set temperature of the heat baths, are placed in
the two ends of the graphene and the temperature difference is denoted as $%
\Delta T=T_{L}-T_{R}$. For the convenience of comparison, here we set $%
T_{L}=350K,T_{R}=300K$, which are near the room temperature. In order to
avoid the spurious global rotation of the graphene in the simulation, we use
fixed boundary condition in the two ends of the graphene. The fixed region
and the heat baths occupy one layer and six layers of atoms respectively. $N$
is the number of the atoms in the heat baths, $k_{B}$ is the Boltzmann
constant and $m$ is the mass of the carbon atom.

We integrate these equations of motion by a Verlet method \cite{Rafii}. The
time step is $0.55$ $fs$, and the simulation runs for $1\times 10^{8}$ time
steps giving a total molecular dynamics time of $55$ $ns$. Generally, $T(t)$
can stabilize around the set value $T_{0}$ after 2 ns for single-layer
graphene and 20 ns for four-layer graphene. Time averaging of the
temperature and heat current is performed from 35 to 55 ns. The heat bath
acts on the particle with a force $-\Gamma p_{i}$; thus the power of heat
bath is $-\Gamma p_{i}^{2}/m$, which can also be regarded as the heat flux
come out of the high temperature heat bath and injected into the low
temperature heat bath. The total heat flux injected from the heat bath to
the system can be obtained by $J=\underset{i}{\sum }\left[ -\Gamma
p_{i}^{2}/m_{i}\right] =-3\Gamma Nk_{B}T(t),$where the subscript $i$ runs
over all the particles in the thermostat \cite{GWu2}. The final thermal
conductivity is calculated from the well-known Fourier's law $\kappa
=Jl/(\Delta Twhn)$, in which $l$, $w$, and $h$ (=0.335nm) are the length,
width and thickness of each layer graphene, respectively. $n$ is the number
of the atomic planes.

In Fig. 1, we explicitly observed the decrease of thermal conductivity as
the number of layers changes from 1 to 8. This change implies a crossover
from 2D graphene to 3D graphite. For the single-layer graphene, the
room-temperature thermal conductivity of zigzag nanoribbons is 2276 Wm$^{-1}$%
K$^{-1}$, which is 43\% larger than that of armchair nanoribbons. From the
inset of Fig. 1, one can find that this attenuation relationship is not
qualitatively varied with the graphene ribbon width. Another interesting
phenomenon, that the zigzag graphene loses its superiority in thermal
conductivity as the number of layers increases to 5, predicts that the
chirality dependence of thermal conductivity may disappear for 3D graphite.
Here we have to point out that the superiority in thermal conductivity of
zigzag nanoribbons is only available for the temperature larger than 220K.
When the temperature is less than 220K, as displayed in Fig. 2, we can
observe that armchair graphene has better thermal conductivity than zigzag
graphene. The temperature-dependent thermal conductivity of armchair and
zigzag graphene can be interpreted by the number of phonons and the path for
phonons passing. From the structure of armchair and zigzag graphene, we can
know that the path in armchair graphene ($1.35l)$ is longer than that ($1.15l
$) in zigzag graphene with the same length, where $l$ is the length of
graphene. Similarly, at the same width, the armchair graphene occupies more
areas and particles than the zigzag graphene (the ratio is about 1.15). At
high temperature, longer path means more phonons scattering. At low
temperature, more particles mean more phonons to conduct heat. Therefore,
the thermal conductivity of armchair is larger at low temperature and
smaller at high temperature than that of the zigzag graphene. 
\begin{figure}[htbp]
\begin{center}\includegraphics[width=8cm,height=6cm]{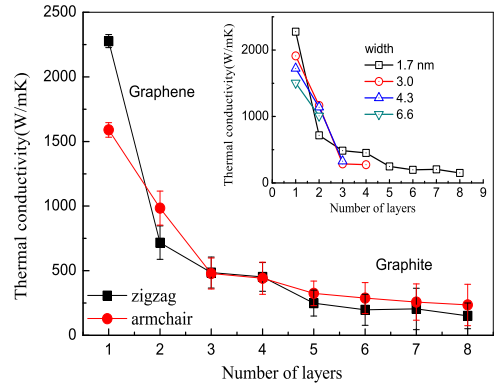}
  \end{center}
  \caption{Layer-dependent thermal conductivity of
few-layer graphene. The 1-layer nanoribbons refer to graphene and the 5%
\symbol{126}8 layers nanoribbons are similar to utra-thin graphite. The
inset is the layer-dependent thermal conductivity for various width zigzag
graphene. The average temperature is 325K.}
   \label{}
\end{figure}

\begin{figure}[htbp]
\begin{center}\includegraphics[width=8cm,height=6cm]{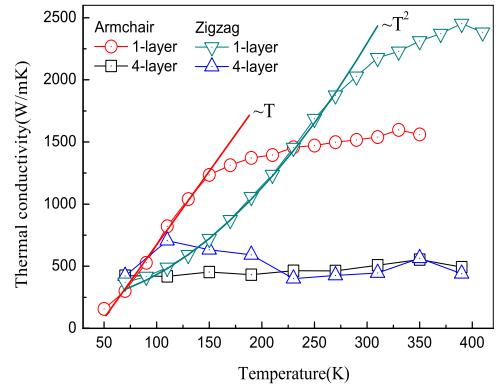}
  \end{center}
  \caption{Temperature
dependence of thermal conductivity of zigzag and armchair graphene for one
and four layers atomic planes.}
   \label{}
\end{figure}

Recent experiments suggest that thermal transport at the nanoscale is
dominated by a ballistic rather than a diffusive mechanism \cite{Ghosh2}.
For the single-layer graphene, figure 3 illustrates that thermal conduction
at low temperature is dominated by the straight or bending mode, with a
power law of \symbol{126}T for armchair graphene and \symbol{126}T$^{2}$ for
zigzag graphene. The former is good agreement with the elastic-shell-based
theoretical results in graphene with finite width \cite{Enrique}. The later
shows that the chirality can change the power law relationship between the
thermal conductivity and the temperature. We explained that the difference
between the power law thermal conductivity of armchair and zigzag graphenes
is mainly due to the different phonon scattering rates at the armchair and
zigzag edges \cite{Jiuning}. This power law relationship implies that the
graphene conduct heat mainly through ballistic transport mode in this
temperature region \cite{Enrique}.

For the multi-layer graphene, the ballistic transport decays at the same
temperature region and the thermal conductivity fluctuates near 500 Wm$^{-1}$%
K$^{-1}$. The crossover from 2D graphene to 3D graphite, which induces the
decrease of thermal conductivity of few-layer graphene, has been explained
as the cross-plane coupling of the low-energy phonons and changes in the
phonon Umklapp scattering \cite{Ghosh}. For the single-layer graphene, i.e.,
in the absence of cross-planes coupling, the thermal transport mode is of
ballistic. Ballistic transport means less collision of phonons, however in
the presence of cross-planes coupling, the phonons will scatters with the
particles at the interface between the layers and then the thermal
conductivity will decrease. Therefore, it is reasonable that the cross-plane
coupling, which is similar to the inter-chain coupling in double
Frenkel-Kontorova chains \cite{Zhong}, will result in a suppression of
thermal conductivity in graphene nanoribbons. Here, from the view of
spatially-resolved phonon spectral behavior shown in Figs. 3(a) and (b), we
present that the phonon occupies a wide frequency region for monolayer
graphene in the absence of layer-to-layer coupling. However, for four-layer
graphene, in the presence of layer-to-layer coupling, as shown in Figs. 3(c)
and (d) the phonons in high frequency region from 1,200 to 1,600 cm$^{-1}$
shrinks and weakens at the edge of nanoribbons. In Refs. \cite{Savin} and
\cite{Shiomi}, the results about carbon nanotubes and graphenes have
illustrated that the high frequency phonon modes, which\ can be affected by
the edge sensitively, play a major role in the low-dimensional non-Fourier
heat conduction. Obviously, the cross-planes coupling can induce unperfect
and rough edges in the graphene. The edge roughness can suppress thermal
conductivity strongly \cite{Savin}. Thus, as the high frequency phonons
shrinks with the layer-to-layer coupling, the material's ability to conduct
heat will decay.

\begin{figure}[htbp]
\begin{center}\includegraphics[width=8cm,height=6cm]{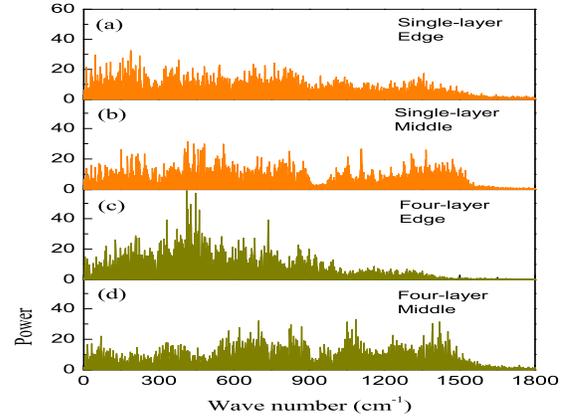}
  \end{center}
  \caption{Spatially-resolved phonon spectrum in the zigzag graphene. Edge (a) and
middle (b) particle in the single-layer graphene, edge (c) and middle (d)
particle in the four-layer graphene. The parameters are $T_{L}=350K,$ $%
T_{R}=300K$.}
   \label{}
\end{figure}

In summary, the thermal conductivity of few-layer graphene strongly depends
on the number of atomic planes. The reduction of the ability of few-layer
graphene to conduct heat is attributed to the crossover from 2D graphene to
3D graphite. We speculated this decreasing evolution in thermal conductivity
is mainly due to the shrinking of high frequency phonon induced by the
cross-layer coupling. We obtained the thermal conductance of zigzag and
armchair nanoribbons as a power law relationship of the temperature at low
temperature. For the multi-layer graphene, the thermal conductivity is
independent on the chirality. Although our calculations indeed show that the
number of layer is disadvantageous to the thermal conductivity in few-layer
graphene, the intriguing questions about the thermal conductivity of
few-layer graphene are expected to stimulate further experimental and
theoretical investigations of phonon transport.

We would like to thank the Siyuan clusters for running part of our programs.
This work was supported in part by the National Natural Science Foundation
of China (Grant No.11004082) and the Natural Science Foundation of Guangdong
Province, China (Grant No.01005249).


\begin{thebibliography}{99}
\bibitem{Lepri} S. Lepri, R. Livi, and A. Politi, Phys. Rep. \textbf{377}, 1
(2003).

\bibitem{Novoselov1} K. S. Novoselov, A. K. Geim, S. V. Morozov, D. Jiang,
Y. Zhang, S. V. Dubonos, I. V. Grigorieva and A. A. Firsov, Science, \textbf{%
306}, 666-669 (2004).

\bibitem{YBZhang} Y. B. Zhang, Y. W. Tan, H. L. Stormer, and P. Kim, Nature,
\textbf{438}, 201 204 (2005).

\bibitem{Nair} R. R. Nair, P. Blake, A. N. Grigorenko, K. S. Novoselov, T.
J. Booth, T. Stauber, N. M. R. Peres and A. K. Geim, Science, \textbf{320},
1308 (2008).

\bibitem{Balandin} A. A. Balandin, S. Ghosh, W. Bao, I. Calizo, D.
Teweldebrhan, F. Miao, C. N. Lau, Nano Lett. \textbf{8}, 902 907 (2008).

\bibitem{WZhang} W. Zhang, Z. Zhu, F. Wang, T. Wang, Litao Sun and Z. Wang,
Nanotechnology, \textbf{15}, 936--939 (2004).

\bibitem{Jiuning} J. Hu, X. Ruan, and Y. P. Chen, Nano Lett., \textbf{9},
2730-2735 (2009).

\bibitem{Ghosh} S. Ghosh, W. Bao, D. L. Nika, S. Subrina, E. P. Pokatilov,
C. N. Lau and A. A. Balandin, Nature Material, \textbf{9,} 555 (2010).

\bibitem{Klemens} P. G. Klemens, J. W. Bandgap Mater. \textbf{7}, 332 339
(2000).

\bibitem{Hochbaum} A. I. Hochbaum, R. Chen, R. D. Delgado, W. Liang, E. C.
Garnett, M. Najarian, A. Majumdar, P. Yang, Nature, \textbf{451}, 163 167
(2008).

\bibitem{GWu1} G. Wu and B. Li, J. Phys.: Condens. Matter, \textbf{20,}
175211 (2008).

\bibitem{Savin} A. V. Savin, Y. S. Kivshar, and B. Hu, Phys. Rev. B, \textbf{%
82}, 195422 (2010).

\bibitem{Chang} C. W. Chang, D. Okawa, A. Majumdar, and A. Zell, Science
314, 1121 (2006).

\bibitem{BWLi} B. Li, Lei Wang, and Giulio Casati, Appl. Phys. Lett. \textbf{%
88}, 143501 (2006).

\bibitem{Tersoff} D. W. Brenner, Phys. Rev. B \textbf{42}, 9458 (1990).

\bibitem{GWu2} G. Wu and B. Li, Phys. Rev. B \textbf{76}, 085424 (2007).

\bibitem{Rafii} H. Rafii-Tabar, Computational Physics of Carbon Nanotubes,
Cambridge University Press, New York (2008).

\bibitem{Ghosh2} S. Ghosh, I. Calizo, D. Teweldebrhan, E. P. Pokatilov, D.
L. Nika, A. A. Balandin, W. Bao, F. Miao, C. N. Lau, Appl. Phys. Lett.,
\textbf{92}, 151911 (2008).

\bibitem{Enrique} E. Munoz, J. Lu, and B. I. Yakobson, Nano Lett., \textbf{10%
}, 1652--1656 (2010).

\bibitem{Zhong} W. R. Zhong, Phys. Rev. E \textbf{81}, 061131 (2010).

\bibitem{Shiomi} J. Shiomi and S. Maruyama, Phys. Rev. B 73, 205420 (2006).
\end{thebibliography}
\end{document}